\documentclass[twocolumn,
showpacs,
]{revtex4}
\usepackage{graphicx}
\begin{document}
\title{Isospin Quantum Number of ${\bf D_{s0}^+(2317)}$}
\author{Arata Hayashigaki}
\affiliation{ Department of Physics, Kyoto University, Kyoto 
606-8502, Japan}
\author{Kunihiko Terasaki}
\affiliation{ Yukawa Institute for Theoretical Physics, 
Kyoto University, Kyoto 606-8502, Japan}
\thispagestyle{empty}
\begin{abstract}
The $D_s^+\pi^0$ and $D_s^{*+}\gamma$ decays of $D_{s0}^{+}(2317)$ 
are studied by assigning it to various charmed strange scalar mesons. 
As a result, it is found that its assignment to an iso-triplet 
four-quark meson is favored by the severest experimental constraint 
on the ratio of the rates for these decays, while assigning it to an 
$I=0$ state (a four-quark or a conventional $\{c\bar{s}\}$) is 
inconsistent with this constraint. 
\end{abstract}
\maketitle
Recently, a narrow $D_s^+\pi^0$ resonance of mass $\simeq 2317$ 
MeV has been observed~\cite{observations,CLEO}, and various proposals 
regarding its assignments have been 
made~\cite{BCL,CH,assignments,Terasaki-D_s}, including the proposal 
that it is the ordinary scalar $D_{s0}^{*+}\sim\{c\bar s\}$~\cite{DGG}, 
which is the chiral partner of $D_s^+$~\cite{chiral}. 
However, its isospin quantum number is not yet known definitely, 
although its neutral and doubly charged partners have not been 
observed~\cite{BABAR04}. This is because the observation of different 
charged states is closely related to their production rates,  
which are still not known. In addition, if it is assigned to 
$D_{s0}^{*+}$,  it is not easy to satisfy the
severest experimental constraint on the ratio~\cite{CLEO}, 
\begin{equation}
R_{\rm exp}\equiv {\Gamma(D_{s0}^{+}(2317)\rightarrow D_s^{*+}\gamma) 
\over 
\Gamma(D_{s0}^{+}(2317)\rightarrow D_s^+\pi^0)}\Biggr|_{\rm exp} < 0.059, 
                                               \label{eq:constraint}
\end{equation}
as discussed in Refs.~\cite{Wang} and ~\cite{Colangelo}. 
In particular, the heavy-hadron chiral perturbation theory~\cite{MS} 
provides a ratio (of leading order) that is larger by at least a factor 
of 2 than the upper bound given in Eq.~(\ref{eq:constraint}), and 
the result of the next-to-leading order includes too many unknown 
parameters to give a definite prediction for this ratio. 
It has also been argued that this ratio would be larger than unity 
if this resonance were assigned to a $DK$ molecule. 
In this article, therefore, we study the 
$D_{s0}^+(2317)\rightarrow D_s^{*+}\gamma$ and 
$D_{s0}^+(2317)\rightarrow D_s^+\pi^0$ decays 
rather phenomenologically (without considering the details of models) 
and demonstrate that Eq.~(\ref{eq:constraint}) can be easily
satisfied if $D_{s0}^{+}(2317)$ is assigned to the iso-triplet scalar 
four-quark meson,  
$\hat F_I^+\sim [cn][\bar s\bar n]_{I=1}\,(n=u,\,d)$, 
while it is difficult to reconcile its assignment to an $I=0$ state 
(the conventional scalar, $D_{s0}^{*+}\sim\{c\bar s\}$, or 
the scalar four-quark, $\hat F_0^+\sim [cn][\bar s\bar n]_{I=0}$) 
with  Eq.~(\ref{eq:constraint}). 
(For notation of four-quark mesons, see Ref.~\cite{Terasaki-D_s}.)  
The possible existence of such four-quark states 
has also been suggested for light flavor mesons 
(the nonet of $[qq][\bar q\bar q]$, $\hat\delta^s$, $\hat\sigma^s$, 
$\hat\sigma$, $\hat\kappa$)~\cite{Jaffe,Achasov,Maiani}, which would 
correspond to the observed light scalar mesons~\cite{PDG04}, 
$a_0(980)$, $f_0(980)$, $\sigma(600)$ and $\kappa(800)$~\cite{E791}, 
respectively. Therefore, we assign the observed scalar nonet to the 
$[qq][\bar q\bar q]$ nonet, so that we can easily determine the 
approximate degeneracy between the observed $a_0(980)$ and $f_0(980)$.  

If $D_{s0}^+(2317)$ is assigned to $\hat F_I^+$, its decay into 
$D_s^+\pi^0$  can proceed through isospin-conserving strong
interactions. 
Therefore, it was naively conjectured that the decay is of the so-called 
{\it fall-apart} type and that it has a high rate ($\sim 100$ MeV or 
more)~\cite{CH}. However, this argument is quite naive, as seen below. 
A four-quark $[qq][\bar q\bar q]$ state can be decomposed into a sum 
of products of $\{q\bar q\}$ pairs with different color and spin, and 
their coefficients are given by the crossing matrices for color and 
spin~\cite{Jaffe}. The $D_s^+$ and $\pi^0$ in the decay 
$\hat F_I^+\rightarrow D_s^+\pi^0$ are color-less and spin-less, and 
the corresponding coefficient is much smaller than unity; i.e., 
the overlap of the wave functions of the initial $\hat F_I^+$ 
and the final $D_s^+\pi^0$ is small. In this way, the narrow width of 
$\hat F_I^+$ can be understood~\cite{Terasaki-ws,Terasaki-ws1} as 
seen more explicitly below. 
The rate for a decay $A \rightarrow BC$ is given by 
\begin{eqnarray}
&&\hspace{-5mm}\Gamma(A \rightarrow BC)          \nonumber\\
&&=\Biggl(\frac{1} {2J_A + 1}\Biggr)
\Biggl(\frac{q_c}{8\pi m_{A}^2}\Biggr)
\sum_{\rm spin}|M(A \rightarrow BC)|^2,  
                                          \label{eq:rate}
\end{eqnarray}
where $J_A$ and $m_A$ denote the spin and mass, respectively, of the 
parent particle $A$, and $q_c$ the center-of-mass momentum of the final 
particle. The amplitude for the decay 
$\hat F_I^+\rightarrow D_s^+\pi^0$, 
as an example, can be approximated as  
\begin{equation}
M(\hat F_I^+ \rightarrow D_s^+\pi^0)_{\rm } 
\simeq \Biggl({m_{\hat F_I}^2 - m_{D_s}^2\over f_{\pi^0}}\Biggr)
\langle{D_s^+|A_{\pi^0}|\hat F_I^+}\rangle                         
                                            \label{eq:tbd-amp}
\end{equation}
by using a hard pion technique with the PCAC (partially conserved 
axial-vector current) in the infinite momentum frame 
(IMF)~\cite{hard-pion,suppl}. 
The {\it asymptotic} matrix element of $A_\pi$ 
(i.e., the matrix element of $A_{\pi}$ between single hadron 
states with infinite momentum), $\langle{B|A_\pi|A}\rangle$, gives 
the dimensionless $AB\pi$ coupling strength. 
Here, we compare $\hat F_I^+\rightarrow D_s^+\pi^0$ with  
$\hat\delta^{s+}\rightarrow \eta\pi^+$ by using {\it asymptotic} 
$SU_f(4)$ symmetry (roughly speaking, the $SU_f(4)$ symmetry of the 
asymptotic matrix elements). The above-mentioned hard pion technique, 
the asymptotic flavor symmetry and their important results are  
reviewed comprehensively in Ref.~\cite{suppl}. 
The amplitude for $\hat\delta^{s+}\rightarrow \eta\pi^+$ is obtained 
by replacing ($\hat F_I^+$, $D_s^+$, $\pi^0$) by ($\hat\delta^{s+}$, 
$\eta$, $\pi^+$) in Eq.~(\ref{eq:tbd-amp}). 
The usual $\eta$-$\eta'$ mixing~\cite{PDG04} leads to 
$\eta = \cos\Theta\cdot\eta^n - \sin\Theta\cdot\eta^s$, 
where $\eta^n$ and $\eta^s$ are the $\{n\bar n\}$ and $\{s\bar s\}$ 
components of $\eta$, respectively, and $\Theta=\chi + \theta_P$, with 
$\cos\chi=1/\sqrt{3}$ and the $\eta$-$\eta'$ mixing angle 
$\theta_P\simeq -20^\circ$, which implies that the $\eta$-$\eta'$ mixing 
is far from the ideal one. 
Because only the $\eta^s$ component survives in the matrix element 
$\langle{\eta|A_{\pi^-}|\hat\delta^{s+}}\rangle$, 
due to the OZI rule, we have
$\langle{\eta|A_{\pi^-}|\hat\delta^{s+}}\rangle
=-\sin\Theta\cdot\langle{\eta^{s}|A_{\pi^-}|\hat\delta^{s+}}\rangle$. 
A {\it naive} application of the asymptotic $SU_f(4)$ symmetry 
yields  
$\sqrt{2}\langle{D_s^+|A_{\pi^0}|\hat F_I^+}\rangle
= \langle{\eta^{s}|A_{\pi^-}|\hat\delta^{s+}}\rangle$. 
This estimation, however, is too naive, since the wave function 
overlapping in $\hat F_I^+\rightarrow D_s^+\pi^0$ is small, as seen 
above, while this is not the case in 
$\hat\delta^{s+}\rightarrow \eta\pi^+$, because 
gluon exchanges in the system, which are more often at a lower energy 
scale, reshuffle color configurations and spin configurations and cause 
configuration mixings. 
As a result, the $\hat\delta^{s+}\rightarrow \eta\pi^+$ decay would 
be nearly of the fall-apart type. 
Therefore, we modify the previously employed asymptotic $SU_f(4)$ 
relation with asymptotic matrix elements (including four-quark mesons) 
by introducing the parameter 
\begin{equation}
\beta =
{\sqrt{2}\langle{D_s^+|A_{\pi^0}|\hat F_I^+}\rangle
\over \langle{\eta^{s}|A_{\pi^-}|\hat\delta^{s+}}\rangle}. 
                                           \label{eq:mod-su(4)}
\end{equation}
In the naive limit of the asymptotic $SU_f(4)$ symmetry, we have 
$\beta=1$, as seen above. We hereafter refer to the symmetry in the 
$\beta\neq 1$ case as the {\it modified} asymptotic $SU_f(4)$ 
symmetry. For simplicity, we consider the limiting case 
in which there is no configuration mixing for $\hat F_I^+$, while 
there is no suppression arising from the crossing matrices for  
$\hat\delta^{s+}$ (and there is maximum mixing of the color and spin 
configurations). In this case, we have $|\beta|^2 = {1\over 12}$ 
from the crossing matrices for color and spin~\cite{Jaffe}.
Inserting 
$|\langle{\eta^s|A_{\pi^-}|\hat \delta^{s+}}\rangle|\sim 0.8$
obtained from the tentative value 
$\Gamma(a_0\rightarrow \eta\pi^+)\simeq 70$ MeV
(noting that the measured rate is $50- 100$ MeV~\cite{PDG04}), we obtain 
\begin{equation}
\Gamma(\hat F_I^+\rightarrow D_s^+\pi^0) \sim 9\,{\rm MeV}.
                                              \label{eq:width-F_I}
\end{equation}
Here $a_0(980)$ has been assigned to $\hat\delta^s$, and 
$|\beta|^2 \simeq {1\over 12}$ has been assumed. This result seems to be 
slightly larger than the measured width of $D_{s0}^+(2317)$~\cite{PDG04} 
but still consistent with the data if the $SU_f(4)$ symmetry breaking 
arising from the overlapping of spatial wavefunctions that was not 
considered above is taken into account. (Note here that the asymptotic 
$SU_f(4)$ symmetry transforming $q\rightarrow c$ overestimates the size 
of the asymptotic matrix elements involving a charmed meson 
state~\cite{Terasaki-ws1} by $\sim 20 - 30\, \%$.) This suggests that 
the situation considered above, in which the strong interactions are 
perturbative at the scale of charm meson mass while 
they are non-perturbative at a scale $\lesssim 1$ GeV, is approximately 
realized, at least in the system under consideration here. In this case, 
the strange quark will be rather slim (with a smaller mass) at the 
scale of the charm meson mass, as will be briefly discussed later. 

When $D_{s0}^+(2317)$ is assigned to an $I=0$ state (the conventional 
scalar $D_{s0}^{*+}$ or the four-quark $\hat{F}_0^+$), the decay 
$D_{s0}^{*+}\rightarrow D_s^+\pi^0\,{\rm or}\,
\hat{F}_0^+ \rightarrow D_s^+\pi^0$ 
does not conserve isospin. As is well known, isospin violating phenomena  
(e.g., the difference between $d$-$u$ quark masses,  
$\Delta m_\pi = m_{\pi^\pm} -m_{\pi^0}$, 
$\Delta m_K = m_{K^0} - m_{K^+}$, etc.) 
have been considered as second-order effects of QED~\cite{Close}, and 
the isospin non-conserving $\pi^0$-$\eta$ mixing~\cite{Dalitz,pi-eta} 
could be of the same origin. (However, this is not assumed in this 
article.) Therefore, the isospin non-conserving decay 
$D_{s0}^{*+}\rightarrow D_s^+\pi^0\,{\rm or}\,
\hat F_0^+ \rightarrow D_s^+\pi^0$, 
which is assumed below to proceed through $\pi^0$-$\eta$ mixing, is 
conjectured intuitively to be (much) weaker than the radiative decay, 
in contrast with the results obtained from specific 
models~\cite{MS,models,AG,BEH}.  

Before studying isospin non-conserving decays of 
$D_s^{*+}$, $D_{s0}^{*+}$ and $\hat F_0^+$, 
we first investigate their radiative decays under the vector meson 
dominance (VMD) hypothesis~\cite{VMD} with the flavor $SU_f(4)$ symmetry 
for the strong vertices and then compare the results with those for the 
$D_s^+\pi^0$ decays. Because the VMD hypothesis with the ideal 
$\omega$-$\phi$ mixing and the flavor $SU_f(3)$ for the strong vertices 
works fairly well for the radiative decays of light vector 
mesons~\cite{photon-mass-dep}, we extend it to a system containing charm 
quarks below. 
The $V\rightarrow P\gamma$ amplitude can be written in the form 
\begin{equation}
M(V\rightarrow P\gamma)       
=\epsilon^{\mu\nu\alpha\beta}G_{\mu\nu}(V)F_{\alpha\beta}(\gamma)
A(V\rightarrow P\gamma),        
                                             \label{eq:V-P-gamma}
\end{equation}
where $V$, $P$ and $A(V\rightarrow P\gamma)$ denote a vector meson, 
a pseudoscalar meson, and the $VP\gamma$ coupling strength, 
respectively, and $G_{\mu\nu}(V)$ and $F_{\alpha\beta}(\gamma)$ are  
the field strengths of a vector meson ($V$) and a photon ($\gamma$), 
respectively. With the VMD hypothesis, $A(V\rightarrow P\gamma)$ 
can be approximated as 
\begin{equation}
A(V\rightarrow P\gamma)
\simeq
\sum_{V'=\rho^0,\,\omega,\,\phi,\,\psi}
\Biggl[{X_{V'}(0)\over m_{V'}^2}\Biggr]A(V\rightarrow PV'),           
                                           \label{eq:V-P-gamma-VMD}
\end{equation}
where $A(V\rightarrow PV')$ denotes the $VPV'$ coupling strength and 
$X_V(0)$ is the $\gamma V$ coupling strength on the photon 
mass shell. (The photon-momentum-square dependence of $X_V$ is 
studied in Ref.~\cite{photon-mass-dep}.) The values of the quantities 
$X_V(0)$ have been estimated from analyses of the photoproduction of 
vector mesons in scattering from various nuclei~\cite{Leith}. 
For $\psi$ photoproduction, both the measured differential cross section 
at $t=0$ and the $\psi N$ total cross section estimated from 
the $A$ dependence of the photoproduction cross sections still have 
large uncertainties. (Here $N$ denotes a nucleon.) 
The necessary values of $X_V(0)$ thus obtained using existing data on 
the photoproduction of vector mesons are as follows: 
$X_\rho(0)=0.033\pm 0.003$ GeV$^2$, 
$X_\omega(0)=0.011\pm 0.001$ GeV$^2$, 
$X_\phi(0)=-0.018\pm 0.004$ GeV$^2$ and
$X_\psi(0)\sim 0.054$ GeV$^2$.) 
[The last value was obtained from 
$d\sigma(\gamma N\rightarrow\psi N)/dt|_{t=0}\simeq 20$ nb/GeV$^2$ 
and
$\sigma_{T}(\psi N) = 3.5 \pm 0.8$ mb~\cite{HLW} 
for the $\psi N$ total cross section.] 
The values of $X_{\phi}(0)$ and $X_{\psi}(0)$ are considerably smaller 
than those of $X_{\phi}(m_\phi^2)$ and $X_{\psi}(m_\psi^2)$ estimated 
from the measured rates for the lepton-pair decays of $\phi$ and 
$\psi$, respectively. Also, the $VPV'$ coupling strength can be 
estimated from the measured rate for the $\omega\rightarrow\pi^0\gamma$ 
decay: Setting $(V,P,V')=(\omega,\pi^0,\rho^0)$ in 
Eq.~(\ref{eq:V-P-gamma-VMD}) and inserting the above value of 
$X_{\rho}(0)$, we obtain 
\begin{equation}
|A(\omega\rightarrow\pi^0\rho^0)|\simeq 18\,{\rm GeV}^{-1}
                                             \label{eq:omega-rho-rho}
\end{equation} 
from 
$\Gamma(\omega\rightarrow \pi^0\gamma)_{\rm exp}=0.734\pm 0.035$ 
MeV~\cite{PDG04}.

To estimate the uncertainties arising from the VMD and the $SU_f(4)$ 
symmetry for $VPV'$ vertices, we focus on the $D^*\rightarrow D\gamma$ 
decays. The amplitudes $A(D^*\rightarrow D\gamma)$ are given by 
the $\rho^0$, $\omega$ and $\psi$-meson poles. Using the $SU_f(4)$ 
relations, 
$2A(D^{*0}\rightarrow D^0\rho^0)
= 2A(D^{*0}\rightarrow D^0\omega) 
= \sqrt{2}A(D^{*0}\rightarrow D^0\psi)
=-2A(D^{*+}\rightarrow D^+\rho^0)         
=2A(D^{*+}\rightarrow D^+\omega)
= \sqrt{2}A(D^{*+}\rightarrow D^+\psi)
=A(\omega\rightarrow \pi^0\rho^0)$, 
and Eq.~(\ref{eq:omega-rho-rho}), we obtain 
$\Gamma(D^{*+}\rightarrow D^+\gamma)\sim 2.4$ keV 
and 
$\Gamma(D^{*0}\rightarrow D^0\gamma)\sim 19$ keV. 
The former is consistent with 
$\Gamma(D^{*+}\rightarrow D^+\gamma)_{\rm exp}\simeq 1.5$ keV 
(with $\sim 30\,\%$ uncertainty) if a possible reduction by 
$\sim 40 - 60$ \% arising from asymptotic $SU_f(4)$ symmetry 
breaking~\cite{Terasaki-ws1} due to the overlapping of spatial 
wavefunctions is taken into account, as discussed above. 
This should be done because the VMD hypothesis with the $SU_f(4)$ 
symmetric $VPV'$ couplings again overestimates the rates for the 
radiative decays of charm mesons, here by $\sim 40 - 60$ \%.  
The latter is consistent with 
$\Gamma(D^{*0}\rightarrow D^0\gamma)_{\rm exp} < 800$ 
keV~\cite{PDG04}. 
The large difference between the above two rates 
results from the fact that the contributions of the $\rho^0$, $\omega$ 
and $\psi$ meson poles interfere constructively in the latter decay 
and destructively in the former. 

Now we are in a position to consider the 
$D_s^{*+}\rightarrow D_s^+\gamma$ decay.  We compare the result with 
that for $D_s^{*+}\rightarrow D_s^+\pi^0$ later. 
The amplitude is given by $\phi$ and $\psi$-meson poles. Using the 
$SU_f(4)$ symmetry relations,  
$\sqrt{2}A(D_s^{*+}\rightarrow D_s^+\phi)
=\sqrt{2}A(D_s^{*+}\rightarrow D_s^+\psi)
=A(\omega\rightarrow \pi^0\rho^0)$, 
and Eq.~(\ref{eq:omega-rho-rho}), we obtain a small rate, 
\begin{equation}
\Gamma(D_s^{*+}\rightarrow D_s^+\gamma)
\sim 0.8 \,{\rm keV},  
                                             \label{eq:VMD-D_s-gamma}
\end{equation}
because of the destructive interference between the $\phi$ and $\psi$ 
pole amplitudes. However, this result satisfies the experimental 
constraint, 
$\Gamma(D_s^{*+}\rightarrow D_s^+\gamma)_{\rm exp}
< 1.8\,\,{\rm MeV}$~\cite{PDG04}. 

Similarly, the amplitude for the radiative decay of the scalar meson 
$S\rightarrow V\gamma$ can be written in the form 
\begin{equation}
M(S\rightarrow V\gamma)       
=G_{\mu\nu}(V)F^{\mu\nu}(\gamma)
                      A(S\rightarrow V\gamma),          
                            \label{eq:S-V-gamma}
\end{equation}
where $S=\hat F_I^+,\,\hat F_0^+$ or $D_{s0}^{*+}$. 
Under the VMD hypothesis, the above $A(S\rightarrow V\gamma)$ is given 
by 
\begin{equation}
A(S\rightarrow V\gamma)
\simeq
\sum_{V'=\rho^0,\,\omega,\,\phi,\,\psi}
\Biggl[{X_{V'}(0)\over m_{V'}^2}\Biggr]
A(S\rightarrow VV').            
                                 \label{eq:S-V-gamma-VMD}
\end{equation}
The $SU_f(4)$ relations relevant to $A(S\rightarrow VV')$ are 
\begin{eqnarray}
&&\hspace{-5mm} A(\hat F_I^{+}\rightarrow D_s^{*+}\rho^0)
=A(\hat F_0^{+}\rightarrow D_s^{*+}\omega)          \nonumber\\
&& 
= A(\phi \rightarrow \hat\delta^{s0}\rho^0)\beta' ,
                                        \label{eq:SU_f(4)-1}\\
&&\hspace{-5mm}2A(D_{s0}^{*+}\rightarrow D_s^{*+}\phi)
=2A(D_{s0}^{*+}\rightarrow D_s^{*+}\psi)            \nonumber\\
&&
=A(\chi_{c0}\rightarrow\psi\psi),    \label{eq:SU_f(4)-2}
\end{eqnarray}
where $\beta'$ denotes a suppression factor (like $\beta$ used above) 
arising from the overlapping of the color and spin wave functions. 
We again assume that the system under consideration is not very far 
from the limiting case, i.e., that of full suppression for 
$A(\hat F_I^{+}\rightarrow D_s^{*+}\rho^0)$ and no suppression for 
$A(\phi\rightarrow\hat\delta^{s0}\rho^0)$ (the so-called fall-apart 
decay). However, the two pairs of quarks $\{q\bar q\}$ are now 
colorless vector states, i.e., $|\beta'|^2 \simeq {1\over 4}$ from 
the crossing matrices for color and spin~\cite{Jaffe}. 
For (i) $S=\hat F_I^+$, (ii) $S=\hat F_0^+$ and (iii) $S=D_{s0}^{*+}$, 
the amplitude $A(S\rightarrow V\gamma)$ is dominated by (i) $\rho^0$, 
(ii) $\omega$, and (iii) $\phi$ and $\psi$ poles, respectively, 
because of the OZI rule. The input data are 
$|A(\phi\rightarrow\hat\delta^{s0}\rho^0)|\simeq 0.020 \,
                                              {\rm MeV}^{-1}$, 
obtained from 
$\Gamma(\phi\rightarrow a_0(980)\gamma)_{\rm exp}=0.32 \pm 0.03$ 
keV~\cite{PDG04}, for (i) and (ii), and 
$|A(\chi_{c0}\rightarrow\psi\psi)|\simeq 0.0823$ MeV$^{-1}$, 
obtained from 
$\Gamma(\chi_{c0}\rightarrow\psi\gamma)_{\rm exp}=119\pm 15\,$ 
keV~\cite{PDG04}, for (iii). 
Inserting Eqs.~(\ref{eq:S-V-gamma-VMD}) and (\ref{eq:SU_f(4)-1}) or 
Eqs.~(\ref{eq:S-V-gamma-VMD}) and (\ref{eq:SU_f(4)-2}) into the 
amplitude, Eq.~(\ref{eq:S-V-gamma}), and taking the values of the 
quantities $X_V(0)$ listed above, we estimate the rates for the 
radiative decays of the scalar mesons listed in Table~I. 
The rate for $\hat F_0^+\rightarrow D_s^{*+}\gamma$ is smaller by an 
order of magnitude than that for $\hat F_I^+\rightarrow D_s^{*+}\gamma$, 
because of the relation $3X_{\omega}(0)\simeq X_{\rho}(0)$. 
\begin{center}
\begin{quote}
{Table~I. Estimated rates for the radiative decays of charmed scalar 
mesons.}
\end{quote}
\vspace{3mm}

\begin{tabular}{c| c| c |c}

\hline
Parent($S$) &\quad$\hat F_I^+\quad $  &\quad $\hat F_0^+\quad $  
&\quad$D_{s0}^{*+}\quad $  \\
\hline
Pole(s) & $\rho^0$ & $\omega$ & $\phi$, $\psi$ \\
\hline
$\Gamma(S\rightarrow D_s^+\gamma)$ 
(keV)
& $\simeq 45$ & $\simeq 4.7$ & $\sim 35$\\
\hline
\end{tabular}
\end{center}
\vspace{2mm}
Therefore, using Eq.~(\ref{eq:width-F_I}) and the results listed in 
Table~I, the ratio reads 
\begin{equation}
R_{\hat F_I}\equiv {\Gamma(\hat F_I^{+}\rightarrow D_s^{*+}\gamma)
\over \Gamma(\hat F_I^{+}\rightarrow D_s^{+}\pi^0)}
\sim 0.005, 
\label{eq:result}
\end{equation}
which satisfies the constraint given in  Eq.~(\ref{eq:constraint}), as 
expected. 

In the above, we have studied the radiative decays of $\hat F_I^{+}$ 
and $\hat F_0^{+}$ by taking only
$\Gamma(\phi\rightarrow a_0(980)\gamma)_{\rm exp}=0.32 \pm 0.03$ 
keV 
as the input data, although another radiative decay, 
$\phi\rightarrow f_0(980)\gamma$, has also been measured. 
The reason for this is that the latter is apparently more uncertain 
than the former. If the measured rate~\cite{PDG04} 
$\Gamma(\phi\rightarrow f_0(980)\gamma)_{\rm exp}=1.87 \pm 0.09$ 
keV 
were taken as the input data, the rate for the 
$f_0(980)\rightarrow \gamma\gamma$ under the VMD hypothesis would be 
given by 
$\Gamma(f_0(980)\rightarrow \gamma\gamma)_{VMD}\simeq 15$ keV, 
which is much larger than  
$\Gamma(f_0(980)\rightarrow \gamma\gamma)_{\rm exp}
=0.39^{+0.10}_{-0.13}$ keV~\cite{PDG04}.
Also, the ratio $R_{\hat F_I}$ would be obtained as 
$R_{\hat F_I}\sim 0.2$ in the same way as in Eq.~(\ref{eq:result}).  
This implies that the electromagnetic interaction would not be very 
much weaker than in the case of ordinary strong interactions, and 
thus it leads to an unnatural result. 
The above would imply that the input data 
$\Gamma(\phi\rightarrow f_0(980)\gamma)_{\rm exp}$ 
is too large. By contrast, if we take  
$\Gamma(\phi\rightarrow a_0(980)\gamma)_{\rm exp}$, 
as before, we can accurately reproduce  the above 
$\Gamma(f_0(980)\rightarrow\gamma\gamma)_{\rm exp}$. 

Next, we discuss isospin non-conserving decays. The amplitude for 
$D_s^{*+}\rightarrow D_s^+\pi^0$ is approximately given by 
\begin{equation}
M(D_s^{*+} \rightarrow D_s^+\pi^0)
\simeq \Biggl({m_{D_s^*}^2 - m_{D_s}^2\over f_{\pi^0}}\Biggr)
\langle{D_s^+|A_{\pi^0}|D_s^{*+}}\rangle,                         
                                         \label{eq:I-non-cons-D_s^*}
\end{equation}
as in Eq.~(\ref{eq:tbd-amp}). Here we have assumed that the violation 
of isospin conservation is caused by the $\eta$-$\pi^0$ mixing, i.e., 
$A_{\pi^0}\rightarrow A_{\pi^0} + \epsilon A_\eta$,
($|\epsilon|\ll 1$), and therefore we replace 
$\langle{D_s^+|A_{\pi^0}|D_s^{*+}}\rangle$ by 
$\epsilon \langle{D_s^+|A_{\eta}|D_s^{*+}}\rangle$. 
The axial charge $A_\eta$ is given by 
$A_{\eta} = \cos\Theta\cdot A_{\eta^n} 
                    - \sin\Theta\cdot A_{\eta^s}$,  
where $A_{\eta^n}$ and $A_{\eta^s}$ are the $\{n\bar n\}$ and 
$\{s\bar s\}$ components of $A_{\eta}$, respectively. 
In the asymptotic matrix element of $A_{\eta}$, only the matrix element 
of $A_{\eta^s}$ can survive, due to the OZI rule, and the asymptotic 
$SU_f(4)$ symmetry leads to 
$\langle{D_s^+|A_{\eta^s}|D_s^{*+}}\rangle 
= {1\over 2}\langle{\pi^+|A_{\pi^+}|\rho^0}\rangle$.
Therefore, it follows that  
\begin{equation}
2\langle{D_s^+|A_{\pi^0}|D_s^{*+}}\rangle 
= -\epsilon\sin\Theta\cdot\langle{\pi^+|A_{\pi^+}|\rho^0}\rangle. 
                                             \label{eq:asymp-eta^s}
\end{equation}
The size of $\langle{\pi^+|A_{\pi^+}|\rho^0}\rangle$ can be estimated as 
$|\langle{\pi^+|A_{\pi^+}|\rho^0}\rangle|\simeq 1.0$~\cite{hard-pion} 
from the measured rate, 
$\Gamma(\rho\rightarrow\pi\pi)_{\rm exp}\simeq 150$ MeV~\cite{PDG04}. 
Inserting Eq.~(\ref{eq:asymp-eta^s}) with $\Theta\simeq
35^\circ\,(\theta_P\simeq -20^\circ)$, as before, into 
Eq.~(\ref{eq:I-non-cons-D_s^*}) and using the measured branching ratios 
for the 
$D_s^{*+}\rightarrow D_s^+\pi^0$ and $D_s^{*+}\rightarrow D_s^+\gamma$ 
decays~\cite{PDG04}, we get 
\begin{equation}
|\epsilon|\sim 1.0\times 10^{-2},        \label{eq:epsilon}
\end{equation}
where we have used Eq.~(\ref{eq:VMD-D_s-gamma}). This value of 
$|\epsilon|$ is consistent with the result given in Ref.~\cite{Dalitz} 
and with the second-order effect of QED, as argued qualitatively above. 

With the help of the above value of $|\epsilon|$, we finally consider 
two cases of the isospin non-conserving decay, 
$D_{s0}^+(2317)\rightarrow D_s^+\pi^0$: 
(a) assuming $D_{s0}^+(2317)$ is the scalar four-quark 
$\hat F_0^+ \sim [cn][\bar s\bar n]_{I=0}$, and  
(b) assuming $D_{s0}^+(2317)$ is the conventional scalar 
$D_{s0}^{*+}\sim \{c\bar s\}$. 
The amplitude for the 
$S^+\,(=\hat F_0^+\,{\rm or}\,D_{s0}^{*+})\rightarrow D_s^+\pi^0$ 
decay can be approximated as  
\begin{equation}
M(S^+\rightarrow D_s^+\pi^0)
\simeq \Biggl({m_{S}^2 - m_{D_s}^2 \over f_{\pi^0}}\Biggr)
               \langle{D_s^+|A_{\pi^0}|S^+}\rangle,
                                   \label{eq:I-spin-viol-S}
\end{equation}
as in Eq.~(\ref{eq:tbd-amp}). Because such a decay is assumed to proceed 
through the $\pi^0$-$\eta$ mixing, as discussed above, we replace the 
matrix elements 
$\langle{D_s^+|A_{\pi^0}|\hat F_0^+}\rangle$ 
and 
$\langle{D_s^+|A_{\pi^0}|D_{s0}^{*+}}\rangle$  
by the OZI-allowed matrix elements 
$\epsilon\cos\Theta\cdot\langle{D_s^+|A_{\eta^n}|\hat F_0^+}\rangle$ 
and 
$-\epsilon\sin\Theta\cdot\langle{D_s^+|A_{\eta^s}|D_{s0}^{*+}}\rangle$,  
respectively. 

In the case (a), the modified asymptotic $SU_f(4)$ symmetry leads to 
$2\langle{D_s^+|A_{\eta^n}|\hat F_0^+}\rangle
= \langle{\pi^+|A_{\eta^s}|\hat \delta^{s+}}\rangle\beta$. 
Then, taking $|\beta|^2\simeq {1\over 12}$ and 
$|\sqrt{2}\langle{\pi^+|A_{\eta^s}|\hat \delta^{s+}}\rangle|
=|\langle{\eta^s|A_{\pi^-}|\hat \delta^{s+}}\rangle|
\sim 0.8$,   
estimated above, we obtain 
\begin{equation}
\Gamma(\hat F_0^+\rightarrow D_s^+\pi^0) \sim 0.7\,\,  {\rm keV}, 
\label{eq:I-viol-hatF_0}
\end{equation}
which is again smaller than 
$\Gamma(\hat F_0^+\rightarrow D_s^{*+}\gamma) \simeq 4.7\,\,{\rm keV}$, 
listed in Table~I. However, the ratio of the decay rates, 
\begin{equation}
R_{\hat F_0}\equiv{\Gamma(\hat F_0^+\rightarrow D_s^{*+}\gamma)
\over \Gamma(\hat F_0^+\rightarrow D_s^+\pi^0)} \sim 7, 
\label{eq:R-zero}
\end{equation}
is still much larger than the experimental upper bound appearing in 
Eq.~(\ref{eq:constraint}), although it is much smaller than that 
conjectured intuitively.  
This is because the radiative $\hat F_0^+\rightarrow D_s^{*+}\gamma$ 
decay is suppressed due to the small $\gamma\omega$ coupling.   

In the case (b), the asymptotic $SU_f(4)$ symmetry leads to 
$\langle{D_s^+|A_{\eta^s}|D_{s0}^{*+}}\rangle 
=\langle{K^+|A_{\pi^+}|K_0^{*0}}\rangle$. 
The size of the right-hand side is estimated to be 
$|\langle{K^+|A_{\pi^+}|K_0^{*0}}\rangle|\simeq 0.29$ 
from the experimental result 
$\Gamma(K_0^{*0}\rightarrow K^+\pi^-)_{\rm exp} = 182 \pm 24$ 
MeV~\cite{PDG04}. 
Here we have assumed that $K_0^{*}(1430)$ is the conventional 
$^3P_0\,\{d\bar s\}$ state~\cite{PDG04}.  
Using the above result for the asymptotic matrix elements, 
the value of $|\epsilon|$ in Eq.~(\ref{eq:epsilon}), 
and $\theta_P=-20^\circ$, we have
\begin{equation}
\Gamma(D_{s0}^{*+}\rightarrow D_s^+\pi^0) \sim 0.6\,\,{\rm keV},
\label{eq:I-viol-F_0}  
\end{equation}
which is much smaller than the other 
estimates~\cite{Colangelo,MS,models,AG,BEH,Harada}. 
In this case, the ratio of the rate for the radiative decay (listed in 
Table~I) to 
the isospin non-conserving rate is given by 
\begin{equation}
R_{D_{s0}^*}\equiv{\Gamma(D_{s0}^{*+}\rightarrow D_s^{*+}\gamma)
\over \Gamma(D_{s0}^*\rightarrow D_s^+\pi^0)} \sim 60, 
\label{eq:R-conv}
\end{equation}
as conjectured intuitively. This value is again much larger than the 
experimental upper limit given in Eq.~(\ref{eq:constraint}). 
It is quite natural that $R_{\hat F_0}$ and $R_{D_{s0}^*}$ in 
Eqs.~(\ref{eq:R-zero}) and (\ref{eq:R-conv}) are much larger than unity, 
as the rates for isospin non-conserving decays are proportional to 
$|\epsilon|^2=O(\alpha^2)$, as discussed above, 
while those for the radiative decays are proportional to 
$\alpha$, where $\alpha$ denotes the fine structure constant. 

We now discuss other models~\cite{Colangelo,MS,models,AG,BEH},  in 
which $D_{s0}^+(2317)$ has been assigned to the chiral partner 
of $D_s^+$ (or $^3P_0\,\,\{c\bar s\}$). Although those models predict 
much smaller values of the ratio $R_{D_{s0}^*}$ than our result, 
given in Eq.~(\ref{eq:R-conv}), most of them are still larger 
than the experimental upper bound given in Eq.~(\ref{eq:constraint}). 
In Ref.~\cite{BEH}, for example, the small ratio 
$R_{D_{s0}^*}\simeq 0.08$ is obtained. This is quite near the 
experimental upper bound appearing in Eq.~(\ref{eq:constraint}), but 
it is still a little larger. In this model, the estimated rate for 
the $D_{s0}^*\rightarrow D_s^+\pi^0$ decay is 
$\Gamma(D_{s0}^*\rightarrow D_s^+\pi^0)_{\rm BEH}\simeq 21.5$ keV, 
which is considerably larger than the other estimates 
($\sim 5 - 10$ keV)~\cite{Colangelo,MS,models,Harada}. 
If the value of the unknown parameter needed to obtain this rate is 
employed, the width of the non-strange partner, $D_0^*\sim \{c\bar n\}$, 
of $D_{s0}^*$ is predicted to be around $ 500$ MeV~\cite{BEH}. 
However, this value is much larger than the measured widths 
($\sim 240 -280$ MeV) of the broad bumps that have been observed in 
the $D^0\pi^+$ and $D^+\pi^-$ invariant mass 
distributions~\cite{BELLE-D_0,FOCUS} and have been interpreted as peaks 
arising from the scalar resonances $D_0^*$. 
In addition, the widths of the resonances $D_0^*$ might be narrower. 
Specifically, the argument~\cite{Terasaki-ws1} has been given that 
these bumps have a structure containing two scalar resonances; 
one is the conventional $D_0^*$ with a width $\sim 50 - 90$ MeV, and 
the other is the four-quark $\hat D\sim [cn][\bar u\bar d]$ with 
a width $\sim 7 - 15$ MeV. Therefore, the value  
$\Gamma(D_{s0}^*\rightarrow D_s^+\pi^0)\simeq 21.5$ keV 
seems to be too large. 
For the radiative $D_{s0}^{*+}\rightarrow D_s^{*+}\gamma$ decay, 
the model~\cite{BEH} predicts 
$\Gamma(D_{s0}^{*+}\rightarrow D_s^{*+}\gamma)_{\rm BEH}\simeq 1.74$ 
keV, 
which is not very far from the other 
predictions~\cite{Colangelo,MS,models,AG}. 
In the model, however, the rate for the radiative decay is very 
sensitive to the assumed values of the constituent quark masses. 
The above small rate has been obtained by choosing $m_c\simeq 3m_s$, 
which is very far from the picture of the heavy charm quark. 
In addition, there is an argument based on a semi-relativistic 
potential model~\cite{MM} that $m_c\simeq 1.4 - 1.5$ GeV and 
$m_s\simeq 0.090 - 0.095$ GeV near the scale of the charm meson mass 
under consideration. Such a small value of $m_s$ might be supported 
by recent lattice QCD analyses~\cite{Gupta}. 
Moreover, the model predicts 
\begin{equation}
{\Gamma(D_s^{*+}\rightarrow D_s^+\pi^0 )
\over 
\Gamma(D_s^{*+}\rightarrow D_s^+\gamma)}\Biggr|_{\rm BEH}
\simeq 0.018,                           \label{eq:R-Ds*-BEH}
\end{equation}
in the case that parameter values yielding  
$\Gamma(D_{s0}^{*+}\rightarrow D_s^{*+}\gamma)_{\rm BEH}\simeq 1.74$ 
keV 
are used.  The above ratio is much smaller than the the well-known 
experimental value~\cite{PDG04}, 
\begin{equation}
{\Gamma(D_s^{*+}\rightarrow D_s^+\pi^0 )
\over 
\Gamma(D_s^{*+}\rightarrow D_s^+\gamma)}\Biggr|_{\rm exp}
=0.062\pm 0.026.                 \label{eq:R-Ds*-exp}
\end{equation}
The model studied in Ref.~\cite{AG} predicts a small rate for the 
$D_{s0}^{*+}\rightarrow D_s^{*+}\gamma$ decay that is consistent with 
the result discussed above.  It also predicts a large rate, 
\begin{equation}
\Gamma(D_{s0}^{*+}\rightarrow D_{s}^+\pi^0) \simeq 129 \,\,{\rm keV},
\end{equation}
in the case that a large value is used for the isospin violating 
parameter, $\epsilon_{AG} \simeq 4.0\times 10^{-2}$, 
which has been estimated by using the measured branching ratios for 
the OZI violating decays~\cite{PDG04}, 
$B(J/\psi(2S)\rightarrow J/\psi(1S)\pi^0)=(9.6\pm 2.1)\times 10^{-4}$ 
and 
$B(J/\psi(2S)\rightarrow J/\psi(1S)\eta)=(3.17\pm 0.21)\times 10^{-2}$. 
As a consequence, a small ratio, $R_{D_{s0}^*} \leq  0.015$, which 
satisfies Eq.~(\ref{eq:constraint}) has been obtained. 
However, the above value of $\epsilon_{AG} $ is much larger than 
the value of $|\epsilon|$ that we estimated from 
the measured ratio in Eq.~(\ref{eq:R-Ds*-exp}). 
This seems to imply that the mechanism of the isospin violation 
discussed above is different from the usual $\pi^0$-$\eta$ mixing. 
For this reason, the model studied in Ref.~\cite{AG} cannot reproduce 
Eq.~(\ref{eq:R-Ds*-exp}). 
We conclude that it is difficult to satisfy the experimental constraint 
given in Eq.~(\ref{eq:constraint}) while simultaneously realizing 
consistency with the well-known ratio Eq.~(\ref{eq:R-Ds*-exp}) 
if $D_{s0}^+(2317)$ is assigned to an $I=0$ state. 

In summary, we have studied the 
$D_{s0}^+(2317)\rightarrow \ D_s^{*+}\gamma$ 
and $D_{s0}^+(2317)\rightarrow D_s^+\pi^0$ decays 
using the VMD hypothesis, the hard pion technique in the IMF, and 
the (asymptotic) $SU_f(4)$. We used the measured rates of  
$\Gamma(\chi_{c0}\rightarrow \psi\gamma)_{\rm exp}$ 
and 
$\Gamma(\phi\rightarrow a_{0}(980)\gamma)_{\rm exp}$ 
as the input data, although the assignment of $a_{0}(980)$ to 
the iso-triplet four-quark meson, $\hat\delta^s$, has not yet been  
confirmed; i.e., it might contain some other configuration, 
like a $K\bar K$ molecule~\cite{KK-molecule}. 
We have also checked ambiguities arising from the above approximations 
by comparing our results for $D^*\rightarrow D\gamma$ (and 
$D^*\rightarrow D\pi$ in 
Ref.~\cite{Terasaki-ws1}) with existing experimental data. In this way, 
we have observed the followings. 
(i) The assignment of $D_{s0}^+(2317)$ to the iso-triplet four-quark 
meson, $\hat F_I^+\sim [cn][\bar s\bar n]_{I=1}$, is consistent with 
the experimental constraint, Eq.~(\ref{eq:constraint}), as seen in 
Eq.~(\ref{eq:result}). 
(ii) It is difficult to reconcile the assignment of $D_{s0}^+(2317)$ to 
an iso-singlet meson (the four-quark $\hat F_0^+$ or the conventional 
$D_{s0}^{*+}$) with Eqs.~(\ref{eq:constraint}) and (\ref{eq:R-Ds*-exp}) 
simultaneously. 
This is quite natural, because the rates for the isospin non-conserving 
decays are proportional to $|\epsilon|^2=O(\alpha^2)$, while those for
the radiative decays are proportional to $\alpha$, the fine structure 
constant. Thus, it is seen that the assignment of $D_{s0}^{+}(2317)$ to 
$\hat F_I^+$ is most reasonable. In addition, this assignment is 
compatible with the observation that the estimated branching fraction 
for $B\rightarrow\bar DD_{sJ}^+$ would be consistent with the 
measured value if the $D_{sJ}^+$ are interpreted as four-quark mesons, 
while it would be much larger than the measured one  if the $D_{sJ}^+$ 
are interpreted as the conventional $\{c\bar s\}$~\cite{Wang}. 
The remaining problem is that neutral and doubly charged partners 
of $\hat F_I^+$ have not yet been observed~\cite{BABAR04}. 
To solve this problem, we need to know the production mechanism of 
the $\hat F_I^{0,+,++}$ mesons. Determining this mechanism is left as 
a future project. 
(iii) Finally, the numerical results obtained in this work should not 
be interpreted too strictly, because our theoretical framework may 
result in a total overestimate of $\sim 40 - 60\,\%$ in the rate. 
Even considering this ambiguity, however, the conclusion of the present 
work is not changed.

\section*{Acknowledgements}
The authors would like to thank K.~Abe and K.~Hagiwara of KEK for 
informing them of the status of experiments on the 
isospin of $D_{s0}^+$. One of the authors (K.~T.) would like to thank 
T.~Onogi of YITP for discussions and the members of the nuclear theory 
groups of the Tokyo Institute of Technology and Kyoto University 
for discussions and comments. 
The research of A.~H. is supported by a research fellowship from 
the Department of Physics, Kyoto University. 
This work of K.~T. is supported in part by 
a Grant-in-Aid for Science Research from the Ministry of Education, 
Culture, Sports, Science and Technology of Japan (Nos. 13135101 and  16540243). 


\begin{thebibliography}{99}
\bibitem{photon-mass-dep} K. Terasaki, Lett. Nuovo Cim. {\bf 31} (1981), 
457; Nuovo Cim. {\bf 66A} (1981), 475. 
The data used in these papers now should be updated. 
It also should be noted that the $SU_f(3)$ for the $VPV'$ couplings
might overestimate the rates for the $K^*\rightarrow K\gamma$ decays, 
by $\sim 20$ per cent.
\end{thebibliography}
\end{document}